\documentclass[a4paper,12pt, epsfig]{article}
\usepackage[utf8]{inputenc} 

\def\letter{0}\def\pr{0}
\pdfoutput=1 

\usepackage{epsfig}
\usepackage{epstopdf}
\usepackage{textcomp}
\usepackage{graphicx}
\usepackage{ifthen}
\usepackage{ulem}

\usepackage{amsmath}
\usepackage[psamsfonts]{amssymb}
\usepackage{euscript}

\usepackage{latexsym}
\usepackage[arrow,matrix,curve]{xy}

\usepackage[hypertexnames=false]{hyperref}
\hypersetup{
    colorlinks=true,
    linkcolor=blue,
    filecolor=magenta,   
    citecolor=black,   
    urlcolor=cyan,
    }

\newskip\humongous \humongous=0pt plus 1000pt minus 1000pt

\newif\ifdtup

\def\,{\hspace{-.1cm}}
\def\hsp{,\hspace{.7cm}}

\def\fc#1#2 {\frac{n}{q}#1\frac{n}{q}#2}

\newcommand{\vac}{\ensuremath{|0\rangle}}

\renewcommand{\cos}{\textrm{cos}}
\renewcommand{\sin}{\textrm{sin}}

\renewcommand{\sinh}{\textrm{sinh}}
\renewcommand{\cosh}{\textrm{cosh}}
\renewcommand{\tanh}{\textrm{tanh}}
\newcommand{\sech}{\textrm{sech}}
\newcommand{\csch}{\textrm{csch}}

\renewcommand{\theequation}{\arabic{section}.\arabic{equation}}
\renewcommand{\(}{\begin{equation}}
\renewcommand{\)}{end{equation} \vspace{-.05in}\linebreak}

\newcounter{saveeqn}
\newcounter{savealpheqn}

\newcommand{\alpheqn}{\setcounter{saveeqn}{\value{equation}}%
  \stepcounter{saveeqn}\setcounter{equation}{0}%
  \renewcommand{\theequation}{\mbox{\arabic{section}.\arabic{saveeqn}
\alph{equation}}}
  \renewcommand{\)}{\end{equation}}}
\def\part#1{\frac{\partial}{\partial{#1}}}%
\def\group#1{\refstepcounter{equation}\setcounter{saveeqn}
 {\value{equation}}%
  \label{#1}\setcounter{equation}{0}%
\renewcommand{\theequation}{\mbox{\arabic{section}.\arabic{saveeqn}
\alph{equation}}}
  \renewcommand{\)}{\end{equation}}}
\newcommand{\reseteqn}{\setcounter{equation}{\value{saveeqn}}%
  \renewcommand{\theequation}{\arabic{section}.\arabic{equation}}%
  \renewcommand{\)}{\end{equation}}}

\newcommand{\aalpheqn}{\setcounter{saveeqn}{\value{equation}}%
  \stepcounter{saveeqn}\setcounter{equation}{0}%
  \renewcommand{\theequation}{\mbox{
        \Alph{subsection}.\arabic{saveeqn}\alph{equation}}}
   \renewcommand{\)}{\end{equation}}}
\newcommand{\areseteqn}{\setcounter{equation}{\value{saveeqn}}%
  \renewcommand{\theequation}{\Alph{subsection}.\arabic{equation}}%
  \renewcommand{\)}{\end{equation}}}

\renewcommand{\(}{\begin{equation}}
\renewcommand{\)}{\end{equation}}
\newcommand{\ba}{\begin{eqnarray}}
\newcommand{\ea}{\end{eqnarray}}

\renewcommand{\sl}{{\sqrt{\lambda}}}

\newcommand{\cbp}{\mathop{\vtop{\ialign{##\crcr
   $\hfil\displaystyle{}\hfil$\crcr\noalign{\kern-13pt\nointerlineskip}
   \BIG{)}\hskip0pt\crcr\noalign{\kern3pt}}}}}
\newcommand{\pa}{\mathop{\vtop{\ialign{##\crcr

$\hfil\displaystyle{\oplus}\hfil$\crcr\noalign{\kern+1pt\nointerlineskip
}
   \hspace{.08in}$^{\alpha=0}$\hskip6pt\crcr\noalign{\kern3pt}}}}}
\renewcommand{\hsp}{,\hspace{.3in}}
\newcommand{\p}{^\prime}
\newcommand{\pp}{^{\prime\prime}}

\catcode`\@=11
\def\vereq#1#2{\lower3pt\vbox{\baselineskip1.5pt \lineskip1.5pt
\ialign{$\m@th#1\hfill##\hfil$\crcr#2\crcr\sim\crcr}}}
\catcode`\@=12

\renewcommand{\(}{\begin{equation}}
\renewcommand{\)}{\end{equation}}

\def\cG{{\mathcal{G}}}
\def\cH{{\mathcal{H}}}

\def\pin#1{\int \frac{d#1}{2\pi}}

\def\df{\mathcal{D}_{f}}

\def\epjc#1{{#1}}
\usepackage[dvipsnames]{xcolor}

\def\hui#1{\textcolor{Mulberry}{Hui: #1}}

\newcommand{\beas}{\begin{eqnarray*}}
\newcommand{\eeas}{\end{eqnarray*}}

\newcommand{\bquo}{\begin{quote}}
\newcommand{\enqu}{\end{quote}}

\def\lim#1{\stackrel{\rm{lim}}{{}_{#1}}}

    \newcommand{\g}{\mathfrak g}

\def\ch{{\mathcal{H}}}

\def\gs#1{\g^{(#1)}}

\def\ok#1{\omega_{k_{#1}}}

\def\gt{\tilde{\g}}

\def\XXint#1#2#3{{\setbox0=\hbox{$#1{#2#3}{\int}$}
     \vcenter{\hbox{$#2#3$}}\kern-.5\wd0}}

\newcommand{\beq}{\begin{equation}}
\newcommand{\eeq}{\end{equation}}
\newcommand{\bea}{\begin{eqnarray}}
\newcommand{\eea}{\end{eqnarray}}

\newskip\humongous \humongous=0pt plus 1000pt minus 1000pt

\newif\ifdtup

\catcode`\@=11

\ifthenelse{\equal{\letter}{0}}{ 

\@addtoreset{equation}{section}
\def\theequation{\arabic{section}.\arabic{equation}}

\def\@normalsize{\@setsize\normalsize{15pt}\xiipt\@xiipt
\abovedisplayskip 14pt plus3pt minus3pt%
\belowdisplayskip \abovedisplayskip
\abovedisplayshortskip \z@ plus3pt%
\belowdisplayshortskip 7pt plus3.5pt minus0pt}

\def\small{\@setsize\small{13.6pt}\xipt\@xipt
\abovedisplayskip 13pt plus3pt minus3pt%
\belowdisplayskip \abovedisplayskip
\abovedisplayshortskip \z@ plus3pt%
\belowdisplayshortskip 7pt plus3.5pt minus0pt
\def\@listi{\parsep 4.5pt plus 2pt minus 1pt
      \itemsep \parsep
      \topsep 9pt plus 3pt minus 3pt}}

\relax

\def\section{\@startsection{section}{1}{\z@}{3.5ex plus 1ex minus  .2ex}{2.3ex plus .2ex}{\large\bf}}

\def\thesection{\arabic{section}}
\def\thesubsection{\arabic{section}.\arabic{subsection}}

\def\appendix{\setcounter{section}{0}
 \def\thesection{Appendix \Alph{section}}
 \def\thesubsection{\Alph{section}.\arabic{subsection}}
 \def\theequation{\Alph{section}.\arabic{equation}}}
\renewcommand{\theequation}{\arabic{section}.\arabic{equation}}

}{
\renewcommand{\theequation}{\arabic{equation}}

} 

\begin{document}

\def\thefootnote{\fnsymbol{footnote}}
\def\thetitle{Oscillon Floquet Modes and the Operators that Excite Them}
\def\hui{Hui Liu}
\def\jarahn{Jarah Evslin\footnote{jarah@impcas.ac.cn}}
\def\tom{Tomasz Roma\'nczukiewicz}
\def\yasha{Yakov Shnir}
\def\andrzej{Andrzej Wereszczy\'nski}
\def\piotr{Piotr Ziobro}
\def\bilguun{Bilguun Bayarsaikhan\footnote{ph.bilguun@gmail.com}}
\def\sujoy{Sujoy Mahato\footnote{ sujoy@impcas.ac.cn
}}

\def\affd{Institute  of  Theoretical Physics,  Jagiellonian  University,  Lojasiewicza  11,  Krak\'ow,  Poland}
\def\ucas{University of the Chinese Academy of Sciences, YuQuanLu 19A, Beijing 100049, China}
\def\imp{Institute of Modern Physics, NanChangLu 509, Lanzhou 730000, China}
\def\affc{Yerevan Physics Institute, 2 Alikhanyan Brothers St., Yerevan 0036, Armenia}
\def\affe{BLTP JINR, Joliot–Curie St 6, Dubna, Moscow region, 141980, Russia}


\ifthenelse{\equal{\pr}{1}}{
\title{\thetitle}
\author{\autone}
\author{\auttwo}
\author{\autthree}
\author{\autfour}
\author{\autfive}
\author{\autsix}
\affiliation {\affa}
\affiliation {\affb}
\affiliation {\affc}
\affiliation {\affd}

}{

\begin{center}
{\large {\bf \thetitle}}

\bigskip

\bigskip


{\large \noindent  \bilguun, \jarahn\ and \sujoy}


\vskip.7cm

1) \imp\\
2) \ucas\\

\end{center}

}

\begin{abstract}
\noindent
We present the scattering modes of the oscillon, to the first three orders in the Fodor expansion, in terms of elementary functions.   In the case of relativistic modes, these results are new. We use them to fully decompose the field and also to analytically invert the decomposition.  In quantum field theory, this allows us to show that the corresponding creation and annihilation operators satisfy the usual oscillator algebra, extending to relativistic modes the demonstration that, at a fixed order in the Fodor expansion, a periodic oscillon state exists.


\end{abstract}

\setcounter{footnote}{0}
\renewcommand{\thefootnote}{\arabic{footnote}}


\section{Introduction}

In Ref.~\cite{univ} it was shown that the nonrelativistic perturbations of small-amplitude oscillons are universal.  In the present note, we extend this result to relativistic perturbations and apply it to the construction of the ground state of a quantum oscillon\footnote{More precisely, this state is not an eigenstate of the Hamiltonian, but rather an eigenstate of the Floquet Hamiltonian.  This implies that it is, at fixed order in the small amplitude expansion, invariant under time evolution by one period of the classical oscillon.}.

Long-lived oscillons exist in the electroweak sector of the Standard Model \cite{graham} when one includes sextic Higgs self-interactions \cite{kasia,kasia2} which are consistent with existing experimental constraints \cite{atlas,cms}.  Collisions of such large-amplitude oscillons, which can pass over the sphaleron barrier to the other vacuum \cite{osckink,nick26,danial26}, may be responsible for baryogenesis in the early Universe.

Oscillons with a small amplitude $\epsilon$ also have a number of cosmological applications \cite{app1,app2,app3,app4,app5} because they are often formed in first order phase transitions \cite{khlop} and they enjoy long but finite lifetimes \cite{oscdec}, with the simulations of Ref.~\cite{longosc} suggesting an unbounded lifetime even in two or three dimensions.  In 1+1 dimensions, which will be the subject of the present work\footnote{Perturbations of oscillons in more dimensions are considerably richer \cite{ddim1,ddim2}.}, they have lifetimes of order $e^{-m/\epsilon}$ where $m$ is the mass of the field of which the oscillon is formed.  In this regime, the oscillon solution itself may be found order by order in $\epsilon/m$ using the expansion of Fodor, Forgacs, Horvath and Lukacs \cite{fodor}, which we will simply call the Fodor expansion.  At each order in the expansion, the oscillon solution is written as a function of the dimensionless combination $\epsilon x$.  At any finite order the oscillon is exactly periodic, although nonperturbatively it radiates \cite{bm76,sk87,v87}.

To understand the quantum oscillon, it is necessary to understand the perturbations of the classical oscillon.  As the oscillon is time-dependent, it does not have normal modes $\g(x)e^{-i\omega t}$.  However, as at finite order in $\epsilon/m$ it is periodic, it instead enjoys Floquet modes $\g(x,t)$, which after a period $2\pi/\Omega$ rotate back to themselves up to a phase
\beq
\g(x,t+2\pi/\Omega)=e^{i\nu}\g(x,t).
\eeq
These Floquet modes are not a basis of all perturbations; however, they can be completed into a basis by adding just two more modes, corresponding respectively to an infinitesimal shift in amplitude and also to the broken boost generator \cite{univ,oscpert}.

It is tempting to apply the Fodor expansion in $\epsilon/m$ to the Floquet modes as well.  However, in the case of continuum Floquet modes with wavenumber $k$, one runs into the problem that $k$ can be any real number and so it may have any scaling with respect to $\epsilon$.  For example, when $k/\epsilon$ is of order unity, these are long perturbations of the size of the oscillon itself.  Larger $k$ correspond to short perturbations.  For example, when $k/m$ is of order unity one finds relativistic perturbations and for higher orders, ultrarelativistic perturbations.  

The question then arises of how to extend the Fodor expansion to such short modes.  One may attempt a double expansion in $\epsilon/m$ and $\epsilon/k$.  Such modes contain a factor $e^{-ikx}$ whose exponent is
\beq
i\frac{(\epsilon x)}{(\epsilon/k)}.
\eeq
In other words, our small parameter appears in the denominator of the exponent, and so the perturbation series fails.  \epjc{This problem can be evaded by first factoring out the $e^{-ikx}$ and then expanding.  Such a factorization is possible because the oscillon is reflectionless.  This }
will be our approach in the current note.

In Ref.~\cite{oscpert} the authors adopted a different convention for the expansion.  Instead of truncating the solution at fixed order, they truncated the equations of motion at finite order in $\epsilon$ and then solved them exactly.  The corresponding solutions were given in terms of hypergeometric functions.  However, the result was rather complicated and so it was not even clear that, in the relativistic case, these modes should approach plane waves.  The expansion presented in the present paper, on the other hand, will express the Floquet modes in terms of elementary functions.

We are motivated by two applications.  First, to find quantum corrections to the energy of the oscillon and also amplitudes for various processes, such as the emission of quantum radiation \cite{hertz,vach}, one needs the Fourier transforms of these modes which can be computed analytically using our new expansion.  

The second and more urgent application is as follows.  In Ref.~\cite{quantosc} we showed that, at the leading order in $\epsilon$, the oscillon is stable.  We did this by constructing a state and showing that, after a period, it evolves to itself.  The state was defined to be that annihilated by all Floquet mode annihilation operators.  The existence of the state thus requires that the Floquet mode annihilation operators commute.  However, we were only able to show this for the nonrelativistic modes.  In the case of relativistic modes, it is clear that they should nearly commute as the modes themselves are nearly plane waves.  Nonetheless, we could not estimate the size of the commutators as they could only be evaluated numerically.  Our new analytical form allows us to compute the commutators of these modes at the first three orders in $\epsilon$, showing that they indeed vanish exactly at these orders as required for the existence of our state.

In fact, in this paper we will go further than in Ref.~\cite{quantosc}.  We will also compute the commutators of the Floquet mode creation operators with the annihilation operators, showing that they 
satisfy the usual oscillator algebra.  This is rather unexpected as the decomposition into Floquet modes is defined by the evolution over a period, whereas the construction of creation and annihilation operators from the field $\phi$ and its conjugate momentum $\pi$ relies on the definition of $\pi$ in terms of an instantaneous time derivative.  Indeed, in Floquet theory the evolution over a period and the instantaneous evolution are generated by two distinct Hamiltonians.  One might well have expected this to result in a mismatch in which the wavenumber $k$ is smeared with a width of order $\epsilon$, or a mixing of Floquet modes with frequencies that differ by a multiple of the oscillon frequency.  

However, our treatment in the current paper is somewhat less general than in Refs.~\cite{univ,oscpert}.  We will restrict our attention to models whose leading interaction, when expanding about a ground state, is quartic.  We call these symmetric oscillons.  In Ref.~\cite{oscpert} we also treated those whose leading interaction is cubic, such as the $\phi^4$ double well model.  It would be straightforward to generalize the present work to models including cubic potentials but we have avoided it here so as to arrive as cleanly as possible at the algebra of the operators that create and destroy Floquet modes.  Needless to say, there are many interesting examples beginning with a cubic interaction and so we would encourage the reader to generalize our results to this case.

After a quick review of Fodor oscillons in Sec.~\ref{gensez}, we will decompose the oscillon's continuum perturbations into two regimes.  First, in Sec.~\ref{nrsez}, we will consider nonrelativistic modes, whose wave numbers are much smaller than the mass $m$ of the scalar field.  Next, in Sec.~\ref{cortsez}, we will consider short modes, whose wavelengths are much shorter than the oscillon, whose length $1/\epsilon$ is itself much longer than the inverse mass $1/m$.  These two regimes cover all wavenumbers and have considerable overlap.  In Sec.~\ref{ortsez} we will find that these modes are not orthogonal, but nonetheless there is a dual basis which is orthogonal to the original modes.  We will show that if the Floquet basis is used to construct quantum fields in terms of Floquet modes, then the dual basis can be used to decompose the fields into Floquet creation and annihilation operators.  Our main result will be that these operators, using the conventions in this note, satisfy commuting copies of the usual oscillator algebra despite the fact that distinct Floquet modes have the same Floquet phase $\nu$.

\section{Generalities} \label{gensez}

\subsection{Fodor Oscillons}

Consider a 1+1 dimensional classical field theory of a scalar field $\phi(x)$ with conjugate momentum $\pi(x)$ and a potential $V$, described by the Hamiltonian
\beq
H=\int dx \left[\frac{\pi(x)^2+\partial_x\phi(x)\partial_x\phi(x)}{2}+\frac{V(\sl\phi(x))}{\lambda}\right].
\eeq
Here $\lambda$ is a coupling constant.   In the quantum theory, $\lambda\hbar/m^2$ is taken to be small, but in the classical theory no such condition is necessary.  We shift $\phi$ so that $\phi=0$ is a local minimum of $V$.  For simplicity we will demand that $V$ is symmetric $V(\phi)=V(-\phi)$ although a general $V$ is treated in Refs.~\cite{univ,oscpert}.  

Let $V^{(n)}$ be the $n$th derivative of $V$ evaluated at $\phi=0$.  We will define the mass $m$ to be
\beq
m=\sqrt{V^{(2)}}.
\eeq

If $V^{(4)}<0$ then for every small, nonzero $\epsilon$ with dimensions of mass, there is an oscillon solution of the equations of motion
\begin{equation}
\phi(x,t)=f(x,t)=\frac{4\epsilon}{\sqrt{-\lambda V^{(4)}}}\sech(\epsilon x)\cos(\Omega t)
+O(\epsilon^3/m^3) \label{osc}
\end{equation}
which oscillates with frequency
\beq
\Omega=\sqrt{m^2-\epsilon^2}.
\eeq

\subsection{Dimensional Analysis}

The coupling constant $\lambda$ will not appear in the equations of motion for the modes and so we will not consider it further.  This leaves four dimensionful quantities:  $m$, $k$, $\epsilon$ and $x$. The first three have dimensions of mass, while $x$ has dimensions of inverse mass.  These can be combined into three independent dimensionless parameters.  Of these, $m$ describes the model and $\epsilon$ describes the oscillon solution.  These will be held fixed, and in particular we will be interested in the limit $\epsilon/m\rightarrow 0$, and so we will expand everything using the Fodor expansion in $\epsilon/m$.  On the other hand, $k$ describes the mode, and so we will be interested in all $k$.  $x$ describes the position, and for each $k$ we need to consider all $x$.

\subsection{The Master Equation}

Let us consider a perturbation of the oscillon by a mode $\g_k$
\beq
\phi(x,t)=f(x,t)+\delta \g_k(x,t)
\eeq
where $\g_k$ is taken to be Floquet, meaning
\beq
\g_k(x,t+2\pi/\Omega)=e^{i\nu}\g_k(x,t)
\eeq
for some phase $e^{i\nu}$, with $\nu$ real.  $\delta$ is an infinitesimal dimensionless parameter.

The classical equations of motion imply that, at linear order in $\delta$, the Floquet modes $\g_k(x,t)$ obey the master equation \cite{oscpert}
\beq
\left(\partial_t^2-\partial_x^2+m^2\right)\g_k(x,t)=\epsilon^2\sech^2(\epsilon x)\left[4+2e^{2i\Omega t}+2e^{-2i\Omega t}\right]\g_k(x,t)+O(\epsilon^4) \label{padeq}
\eeq
where $\g_k$ is normalized so as to be of order $O(\epsilon^0)$.  Note that $\lambda$ and $V^{(4)}$ do not appear in the master equation, it is universal \cite{univ}. 

\epjc{In the case of continuum modes, $\g_k$ will asymptotically reduce to a plane wave.  In principle, this may contain both $e^{ikx}$ and $e^{-ikx}$ components.  However, we will see that oscillons are reflectionless.  Therefore, a basis of fluctuations exist which asymptote to $e^{-ikx}$ in both directions.  To derive a consistent $\epsilon$-expansion in the short wavelength regime, it will be necessary to factor out this term
\beq
\g_k(x,t)=\gt_k(\epsilon x,t)e^{-ikx}.
\eeq
}

\epjc{Substituting this into Eq.~(\ref{padeq}) one arrives at
\beq
\left(\partial_t^2-\partial_x^2+2ik\partial_x+\ok{}^2\right)\gt_k(\epsilon x,t)=\epsilon^2\sech^2(\epsilon x)\left[4+2e^{2i\Omega t}+2e^{-2i\Omega t}\right]\gt_k(\epsilon x,t)+O(\epsilon^4) \label{padeqb}
\eeq
where
\beq
\ok{}=\sqrt{m^2+k^2}.\eeq
In terms of the dimensionless combination $\epsilon x$ this may be written
\beq
\left(\partial_t^2-\epsilon^2 \partial_{\epsilon x}^2+2i\epsilon k\partial_{\epsilon x}+\ok{}^2\right)\gt_k(\epsilon x,t)=\epsilon^2\sech^2(\epsilon x)\left[4+2e^{2i\Omega t}+2e^{-2i\Omega t}\right]\gt_k(\epsilon x,t)+O(\epsilon^4) .\label{padeqb}
\eeq
}

More generally, if $V^{(3)}$ were nonvanishing, this equation would depend on $V^{(4)}$ and $V^{(3)}$, however the leading order nonrelativistic perturbations remain universal \cite{univ} while the leading order relativistic perturbations are plane waves.  Therefore, although the master equation would no longer be universal, at leading order the deformations $\g(x,t)$ would nonetheless be universal.

\section{Nonrelativistic Modes} \label{nrsez}

Nonrelativistic modes are those for which $|k/m|\ll 1$, allowing us to perform a power series expansion in $k/m$ at each order in the Fodor expansion in $\epsilon/m$.  This case includes and generalizes the intermediate and short wavelength modes considered in Ref.~\cite{oscpert}, corresponding respectively to the cases in which $k/\sqrt{m\epsilon}$ and $k/\epsilon$ are of order unity.

Proceeding as there, one may expand
\epjc{
\beq
\gt_k(\epsilon x,t)=G_k(\epsilon x)e^{-i\ok{} t}+H_k(\epsilon x)e^{-i(-2\Omega+\ok {})t}+O(\epsilon^2/m^2) \label{ghexp}
\eeq
}
where the $O(\epsilon^2/m^2)$ corrections include those with time-dependence $e^{-i(2\Omega+\ok{})t}$, which unlike the $e^{i(2\Omega-\ok{})t}$ terms do not receive a resonant enhancement at small $k$.

The $e^{-i\ok{}t}$ terms in the master equation (\ref{padeqb}) are then
\epjc{\beq
-\left(\epsilon^2\partial_{\epsilon x}^2-2i\epsilon k\partial_{\epsilon x}\right)G_k(\epsilon x)=\epsilon^2\sech^2(\epsilon x)\left[4G_{k}(\epsilon x)+2H_k(\epsilon x)\right] \label{eqa}
\eeq
}
while the $e^{-i(-2\Omega+\ok{})t}$ terms are
\epjc{
\beq
\left(-(-2\Omega+\ok{})^2-\epsilon^2 \partial_{\epsilon x}^2+2i\epsilon k\partial_{\epsilon x}+\omega_k^2\right)H_k(\epsilon x)=\epsilon^2\sech^2(\epsilon x)\left[4H_{k}(\epsilon x)+2G_k(\epsilon x)\right]. \label{eqb}
\eeq
}

The coefficient on the left hand side of Eq.~(\ref{eqb}) is
\epjc{\beq
C=\omega_k^2-(-2\Omega+\ok{})^2=4\Omega\ok{}-4\Omega^2=\frac{2}{m}(2m^2-\epsilon^2)(\ok{}-m)+2\epsilon^2.
\eeq
}
In the nonrelativistic limit we identify $\ok{}-m$ as the kinetic energy $k^2/(2m)$ and so this simplifies to
\epjc{
\beq
C=k^2\left(2-\frac{\epsilon^2}{m^2}\right)+2\epsilon^2=2(k^2+\epsilon^2)+O\left(k^2\frac{\epsilon^2}{m^2}\right).
\eeq
}
Eq.~(\ref{eqb}) then reduces to
\epjc{\beq
\left(-\epsilon^2 \partial_{\epsilon x}^2+2(i\epsilon k\partial_{\epsilon x}+k^2+\epsilon^2)\right)H_k(\epsilon x)=\epsilon^2\sech^2(\epsilon x)\left[4H_{k}(\epsilon x)+2G_k(\epsilon x)\right].
\eeq
}
Notice that in both Eqs.~(\ref{eqa}) and (\ref{eqb}), the terms on the left hand side of order $O(m^2)$ have canceled, leaving the smaller terms of orders $O(k^2)$ and $O(\epsilon^2)$.  This cancellation is responsible for the resonant enhancement which leaves $H_k/G_k$ larger than the $O(\epsilon^2/m^2)$ which might naively be expected from the appearance of $\epsilon^2$ on the right hand sides, and it does not occur for the $e^{-i(2\Omega+\ok{})t}$ terms that we have dropped.

Despite the greater generality in our range of wavenumbers $k$, these equations are the same as in Refs.~\cite{univ,oscpert} \epjc{with the $e^{ikx}$ terms factored out, }and so the solutions are again
\epjc{
\beq
G_k(\epsilon x)=\left(1-\frac{\epsilon^2}{k^2}+\frac{\epsilon^2}{k^2}\sech^2(\epsilon x)-2i\frac{\epsilon}{k}\tanh(\epsilon x)\right)\hsp
H_k(\epsilon x)=\frac{\epsilon^2}{k^2}\sech^2(\epsilon x). \label{nrfin}
\eeq
}
Note that this is not an expansion in $\epsilon/k$, which may be larger than or smaller than unity.  Instead, it is reliable up to corrections suppressed by powers of $\epsilon^2/m^2$ and $k^2/m^2$, which are both much less than unity.

\section{Short Modes} \label{cortsez}

\subsection{The Expansion}

As $\gt$ is Floquet, it may be expanded in time harmonics
\beq
\gt_k(\epsilon x,t)=\sum_{n=-\infty}^\infty\g^{(n)}_k(\epsilon x)e^{i(n\Omega-\ok{})t}.
\eeq
Substituting this into our master equation (\ref{padeqb}) one finds that the coefficient of $e^{i(n\Omega-\ok{})t}$ is
\bea
nm(2\ok{}-nm)\g_k^{(n)}(\epsilon x)&=&-2i\epsilon k\partial_{\epsilon x}\g_k^{(n)}(\epsilon x)+\epsilon^2\left[\partial_{\epsilon x}^2+ 
n\left(\frac{\ok{}}{m}-n
\right)+4\sech^2(\epsilon x)\right]\g_k^{(n)}(\epsilon x)\nonumber\\&&+2\epsilon^2
\sech^2(\epsilon x)\left(\g_k^{(n-2)}(\epsilon x)+\g_k^{(n+2)}(\epsilon x)\right)+O(\epsilon^4) \label{padc}
\eea
where $O(\epsilon^4)$ corrections arise from our power series expansion of $\Omega$.

Let us now turn our attention to the regime $|k/\epsilon|\gg 1$, corresponding to modes whose wavelength is much smaller than the oscillon itself, whose size is $1/\epsilon$.  Note that this regime, together with the nonrelativistic regime considered in Sec.~\ref{nrsez}, covers all values of $k$ and also the two regimes overlap when $\epsilon\ll |k|\ll m$.

Up to an overall normalization, $\g$ may be written in terms of the dimensionless quantities $\epsilon/k$, $\epsilon/m$ and $\epsilon x$.  In this small wavelength regime we may expand in powers of the small, dimensionless parameters $\epsilon/k$ and $\epsilon/m$.  The most general form of such an expansion would be
\beq
\g_k^{(n)}(\epsilon x)=\sum_{i_1,i_2=0}^\infty \left(\frac{\epsilon}{k}\right)^{i_1} \left(\frac{\epsilon}{m}\right)^{i_2} \g_{i_1i_2,k}^{(n)}(\epsilon x). \label{dexp}
\eeq
Our approach to this double expansion will be to simply expand in powers of the dimensionful parameter $\epsilon$
\beq
\g_k^{(n)}(\epsilon x)=\sum_{i=0}^\infty \epsilon^i \g_{i,k}^{(n)}(\epsilon x).
\eeq
Here $i$ is the sum of the exponents $i_1$ and $i_2$ in the double expansion (\ref{dexp}).  On the other hand, $\epsilon x$ is treated exactly and is not constrained.  
We will insert this expansion into our new master equation (\ref{padc}).  The extremely ultrarelativistic modes $k\gg m^2/\epsilon$ will not be treated here, so that the $k$ and $\ok{}$ in Eq.~(\ref{padc}) do not affect the power counting.  

\subsection{Order $O(\epsilon^0$)}

At order $O(\epsilon^0)$, Eq.~(\ref{padc}) reduces to
\beq
nm(2\ok{}-nm)\g_{0,k}^{(n)}(\epsilon x)=0.
\eeq
The solutions $2\ok{}=nm$ correspond to an $e^{i\ok{}t}$ dependence, and can be obtained from the present solutions by complex conjugation.  Therefore we will not consider this branch further.  We thus conclude that $\g_{0,k}^{(n)}$ vanishes for all nonzero $n$.

\subsection{Order $O(\epsilon^1$)}

At the next order, the $n=0$ case of Eq.~(\ref{padc}) is
\beq
0=-2i k\partial_{\epsilon x}\g_{0,k}^{(0)}(\epsilon x)
\eeq
and so we conclude that $\g_{0,k}^{(0)}$ is a constant.  We will normalize our Floquet modes so that
\beq
\g_{0,k}^{(0)}=1.
\eeq

The $n\neq 0$ cases of Eq.~(\ref{padc}) are
\beq
nm(2\ok{}-nm)\g_{1,k}^{(n)}(\epsilon x)=-2i k\partial_{\epsilon x}\g_{0,k}^{(n)}(\epsilon x)=0
\eeq
and so we conclude that $\g_{1,k}^{(n)}$ also vanishes for all nonzero $n$.

\subsection{Order $O(\epsilon^2$)}

In the case $n=0$, Eq.~(\ref{padc}) now reduces to
\beq
0=-2i k\partial_{\epsilon x}\g_{1,k}^{(0)}(\epsilon x)+\left[\partial_{\epsilon x}^2+4\sech^2(\epsilon x)\right]\g_{0,k}^{(0)}(\epsilon x)=-2i k\partial_{\epsilon x}\g_{1,k}^{(0)}(\epsilon x)+4\sech^2(\epsilon x).
\eeq
This is solved by
\beq
\g_{1,k}^{(0)}(\epsilon x)=-\frac{2i}{k}\tanh(\epsilon x).
\eeq

The $n\neq 0$ cases are
\beq
nm(2\ok{}-nm)\g_{2,k}^{(n)}(\epsilon x)=2 
\sech^2(\epsilon x)\left(\g_{0,k}^{(n-2)}(\epsilon x)+\g_{0,k}^{(n+2)}(\epsilon x)\right)=2 
\sech^2(\epsilon x)\left(\delta_{n,2}+\delta_{n,-2}\right).
\eeq
We conclude that these are nonvanishing only for $n=\pm 2$
\beq
\g_{2,k}^{(-2)}(\epsilon x)=-\frac{\sech^2(\epsilon x)}{2m(\ok{}+m)}\hsp
\g_{2,k}^{(2)}(\epsilon x)=\frac{\sech^2(\epsilon x)}{2m(\ok{}-m)}.
\eeq

\subsection{Order $O(\epsilon^3$)}

The last case that we will consider is the $n=0$ case at order $O(\epsilon^3)$, where Eq.~(\ref{padc}) becomes
\bea
0&=&-2i k\partial_{\epsilon x}\g_{2,k}^{(0)}(\epsilon x)+\left[\partial_{\epsilon x}^2+4\sech^2(\epsilon x)\right]\g_{1,k}^{(0)}(\epsilon x)\\
&=&-2i k\partial_{\epsilon x}\g_{2,k}^{(0)}(\epsilon x)-\frac{4i}{k}\sech^2(\epsilon x)\tanh(\epsilon x).
\nonumber
\eea
This is solved by
\beq
\g_{2,k}^{(0)}(\epsilon x)=\frac{\sech^2(\epsilon x)}{k^2}.
\eeq

\subsection{Conclusions}

Summarizing, we have found that, up to corrections of order $O(\epsilon^3)$, the Floquet modes are
\bea
\gt_k(\epsilon x,t)&=& e^{-i \ok{}t}\left[1+\frac{\epsilon^2}{k^2}\sech^2(\epsilon x)-\frac{2i\epsilon}{ k}\tanh(\epsilon x)
+\frac{\epsilon^2\sech^2(\epsilon x)}{2 m}\left(
\frac{e^{2i\Omega t}}{\ok{}-m}-\frac{e^{-2i\Omega t}}{\ok{}+m}
\right)
\right]\nonumber\\
&&\hspace{-2.5cm}= e^{-i \ok{}t}\left[1+\frac{\epsilon^2}{k^2}\sech^2(\epsilon x)-\frac{2i\epsilon}{ k}\tanh(\epsilon x)
+\frac{\epsilon^2\sech^2(\epsilon x)}{k^2}\left( \cos(2\Omega t)+i\frac{\ok{}}{m}\sin(2\Omega t)
\right)
\right].\label{corte}
\eea
The absence of an $e^{2ikx}$ term confirms that the small amplitude oscillon is reflectionless
up to corrections of order $O(\epsilon^3)$.

\subsection{Comparisons}

In the regime $\epsilon\ll |k|\ll m$, the perturbations are both nonrelativistic and also short, and so one expects the results here to agree with those of Sec.~\ref{nrsez}.  

To compare them, we note that the expansion Eq.~(\ref{ghexp}) is related to the present expansion by
\beq
G_k(\epsilon x)=\alpha\g_k^{(0)}(\epsilon x)\hsp H_k(\epsilon x)=\alpha\g_k^{(2)}(\epsilon x)
\eeq
where $\alpha$ is a constant reflecting the normalization conditions in the two sections.  Comparing the results Eq.~(\ref{nrfin}) and the first line of Eq.~(\ref{corte}), one observes
\beq
\alpha=1-\frac{\epsilon^2}{k^2}.
\eeq
Here, as $k\ll m$, we have approximated the $\ok{}-m$ in the denominator of $\g^{(2)}$ by $k^2/(2m)$.  On the other hand, the denominator of the $\g^{(-2)}$ term is $\ok{}+m$ which is approximately $2m$.  As a result, $\g^{(-2)}/\g^{(2)}$ is of order $k^2/m^2$ which is small in the nonrelativistic limit, and so $\g^{(-2)}$ is not reported in Sec.~\ref{nrsez}.  

In fact, the nonrelativistic result Eq.~(\ref{nrfin}) and the short wavelength result Eq.~(\ref{corte}) appear to agree even in the nonrelativistic regime $|k|<\epsilon$.  This is rather unexpected, as the latter is an expansion in $\epsilon/k$.  Nonetheless, it implies that at the order at which we have worked, Eq.~(\ref{corte}) alone is sufficient to summarize our results for all $|k|\ll m^2/\epsilon$.

\begin{figure}
    \centering   \includegraphics[width=.45\linewidth]{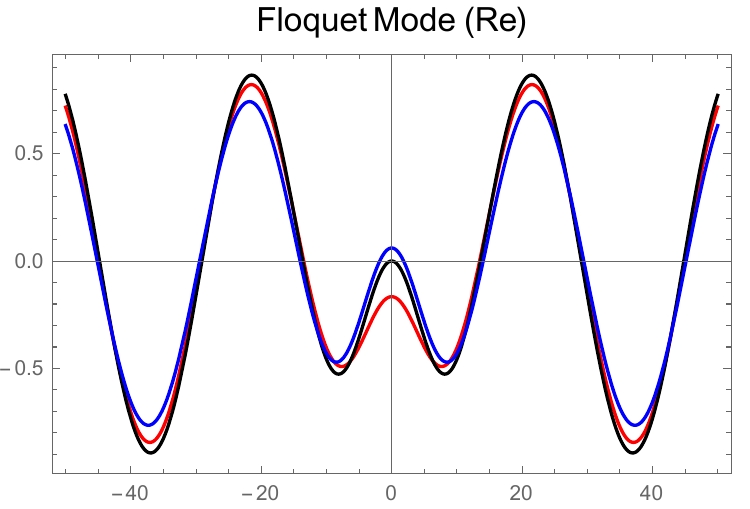}
\includegraphics[width=.45\linewidth]{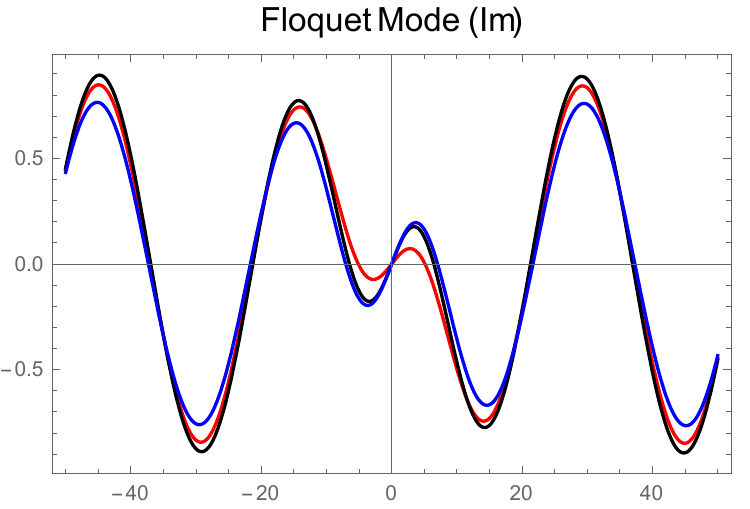}
    \caption{Setting $m=1$ and $\epsilon=0.1$, these are the real (left) and imaginary (right) parts of the $k=0.2$ Floquet modes at time $t=0$, using the expansion of Ref.~\cite{oscpert} (red), the nonrelativistic formula Eq.~(\ref{nrfin}) (black) and the short mode formula Eq.~(\ref{corte}) (blue).  They have been normalized to have amplitude one asymptotically and the leading order plane wave has been subtracted.}
    \label{f2figa}
\end{figure}

\begin{figure}
    \centering   \includegraphics[width=.45\linewidth]{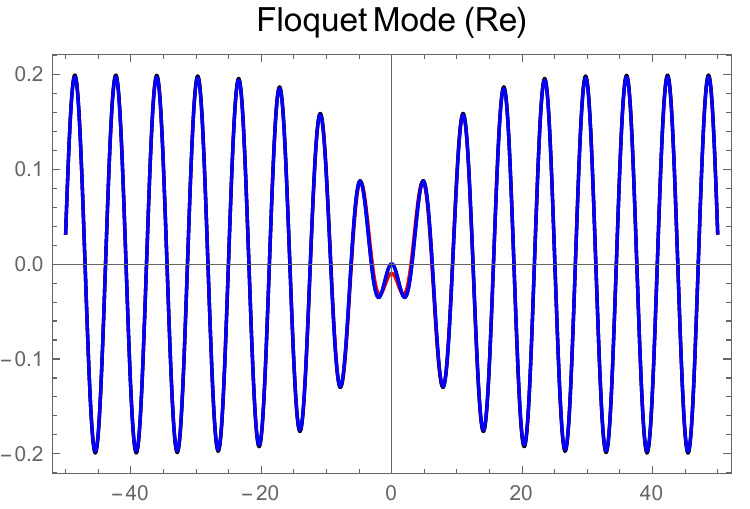}
\includegraphics[width=.45\linewidth]{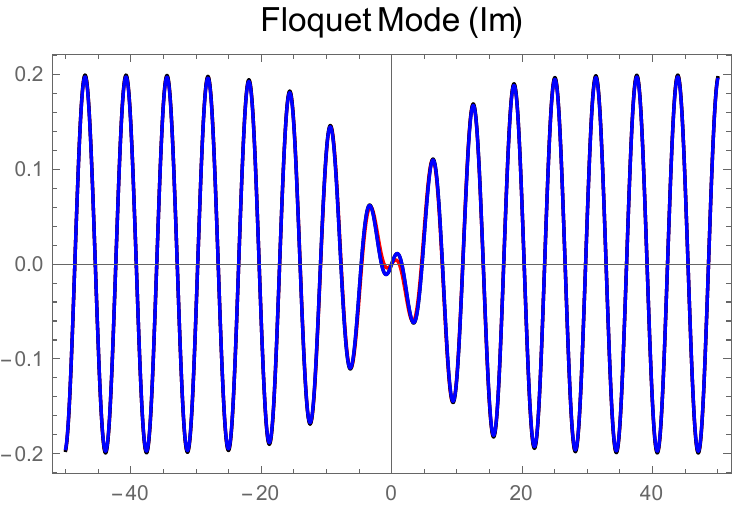}
    \caption{As in Fig.~\ref{f2figa} but with $k=1$.  The corrections to the plane wave are dominated by the phase shift, as it is of order $O(\epsilon/k)$ and the other corrections are of order $O(\epsilon^2/k^2)$.  This phase shift is the same for all three expansions.}
    \label{f1figa}
\end{figure}

In Figs.~\ref{f2figa} and \ref{f1figa} we compare the hypergeometric function corrections to the plane wave computed in Ref.~\cite{oscpert} with the nonrelativistic and short mode approximations $\g_k$ in the present paper.  In general they are quite similar, and are difficult to distinguish once $k/\epsilon$ grows to be of order ten.  The nonrelativistic formula becomes exact at small $k$.  Here, we see that the short mode approximation Eq.~(\ref{corte}) of the present work is closer to it than is the hypergeometric function of Ref.~\cite{oscpert} \epjc{at $|x|\lesssim 5$, where the hypergeometric function exhibits an error in excess of 50 percent}.

\section{Orthogonality} \label{ortsez}

\subsection{Dual Modes}

To decompose quantum fields in terms of these modes, and in particular to invert the decompositions so as to arrive at the commutation relations of the operators that create and destroy the Floquet modes, we need to understand their orthogonality.  In this section, we will describe the orthogonality properties of the short modes Eq.~(\ref{corte}).

Unlike normal modes, whose orthogonality is guaranteed because they solve a Sturm-Liouville problem, Floquet modes are not orthogonal to one another.  Rather, the decompositions may be inverted using their dual modes, which exist at any fixed time.  For example, at time $t=0$, one has the orthogonality relation
\beq
\int dx \g^*_{k_1}(x,0) \g_{D,k_2}(x)=2\pi\delta(k_1-k_2)
\eeq
where the dual mode is
\beq
\g_{D,k}(x)=\frac{k^2}{k^2+4\epsilon^2}e^{-ikx}\left(1-\frac{2i\epsilon}{k}\tanh(\epsilon x)\right)+O(\epsilon^3).
\eeq
This suffices to invert the decomposition of a scalar field $\phi(x)$ via Floquet modes $\g_k(x,0)$ as in Ref.~\cite{quantosc}.

The conjugate momentum $\pi(x)$ is decomposed not using the Floquet modes themselves but rather their time derivatives
\beq
\dot\g_k(x,0)=i\ok{}\left(-\g_k(x,0)+2\frac{\epsilon^2}{k^2}e^{-ikx}\sech^2(\epsilon x)\right)=-i\ok{}\left(1+4\frac{\epsilon^2}{k^2}\right)\g_{D,k}(x)+O(\epsilon^3)\label{gd}
\eeq
whose dual is simply $-ik^2/(\ok{}(k^2+4\epsilon^2))\g_k(x,0)$.

\subsection{The Decomposition}

Let $\hat\phi(x)$ and $\hat\pi(x)$ be two superpositions of short Floquet modes
\beq
\hat\phi(x)=\pin{k}\g_k(x,0) \phi_k\hsp
\hat\pi(x)=i\pin{k}\dot\g_k(x,0) \pi_k. \label{reldec}
\eeq
Note that, up to corrections of order $O(\epsilon^3)$, $\hat\pi(x)$ can be decomposed directly in terms of $\g_{D}$
\beq
\hat\pi(x)=\pin{k}\ok{}\left(1+4\frac{\epsilon^2}{k^2}\right)\g_{D,k}(x)\pi_k. \eeq
One can impose that the Floquet modes are short by letting $\phi_k$ and $\pi_k$ vanish unless $|k/\epsilon|\gg 1$.  

\subsection{Projections}

These decompositions may be inverted by
\beq
\phi_k=\int dx\ \hat\phi(x)\g^*_{D,k}(x)\hsp \pi_k=\frac{1}{\ok{}}\frac{k^2}{k^2+4\epsilon^2}\int dx\ \hat\pi(x)\g^*_{k}(x,0). \label{invdec}
\eeq
While the inversion of $\phi$ is exact as it results from the definition of $\g_D$, that of $\pi$ may have corrections of order $O(\epsilon^3)$.

Now let us turn to quantum field theory.  Let $\phi(x)$ and $\pi(x)$ be operators satisfying canonical commutation relations.  They can only be decomposed using a complete basis of functions, and the relativistic Floquet modes are not complete.  However, the basis can be completed by adding nonrelativistic modes \cite{univ,oscpert} which oscillate over length scales of order $O(1/\epsilon)$, leading to a decomposition
\beq
\phi(x)=\phi_{\rm{NR}}(x)+\pin{k}\g_k(x,0) \phi_k\hsp
\pi(x)=\pi_{\rm{NR}}(x)+i\pin{k}\dot\g_k(x,0) \pi_k
\eeq
where $\phi_{\rm{NR}}(x)$ and $\pi_{\rm{NR}}(x)$ consist of nonrelativistic modes.  When a nonrelativistic mode is multiplied by our relativistic modes and integrated over $x$, the result is of order $e^{-\pi k/\epsilon}$, which vanishes to all orders in $\epsilon$.


As the oscillon is only periodic at finite orders in the $\epsilon$ expansion, such corrections are beyond our analysis.  Therefore, the inverse decompositions Eq.~(\ref{invdec}) continue to hold within our $\epsilon$ expansion.  To see this in a simple example, consider the non-Floquet contribution to $\phi_{\rm{NR}}(x)$ from the mode
\beq
\g_\epsilon(x,t)=\sech(\epsilon x)\cos(\Omega t)
\eeq
that shifts the oscillon's amplitude.  The corresponding contribution to $\phi_{\rm{NR}}$ is \label{quantosc}
\beq
\hat\epsilon \g_\epsilon(x,0)=\hat\epsilon \sech(\epsilon x)
\eeq
where $\hat\epsilon$ is the operator whose eigenvalue is the oscillon's amplitude.  The corresponding contribution to $\phi_k$ in Eq.~(\ref{invdec}), obtained by including this contribution to $\phi(x)$ on the right hand side, is
\bea
\Delta\phi_k&=&\hat\epsilon \int dx \sech(\epsilon x)  \g^*_{D,k}(x)\label{exeq}\\
&=&\frac{k^2}{k^2+4\epsilon^2}\hat\epsilon \int dx \sech(\epsilon x)  
e^{ikx}\left(1+\frac{2i\epsilon}{k}\tanh(\epsilon x)\right)\nonumber\\
&=&\frac{k^2}{k^2+4\epsilon^2}\hat\epsilon\left[ 
\frac{\pi}{\epsilon}+\frac{2i\epsilon}{k}\frac{i\pi k}{\epsilon^2}
\right]\sech{\left(\frac{\pi k}{2\epsilon}\right)}\nonumber\\
&=&-\frac{\pi \hat\epsilon}{\epsilon}\frac{k^2}{k^2+4\epsilon^2}\sech{\left(\frac{\pi k}{2\epsilon}\right)}.\nonumber
\eea
Therefore the contamination of the operator $\hat\epsilon$ into the projection defining $\phi_k$ from $\phi(x)$ is exponentially suppressed in $k/\epsilon$, and so apparently vanishes order by order in our $\epsilon/k$ expansion.  However our computation of the Floquet modes is only reliable at the first three orders, and so one cannot conclude that there is no contamination beyond that.

The projection onto the operator $\hat\epsilon$ in Ref.~\cite{quantosc} was
\beq
\hat\epsilon=\frac{\epsilon}{2}\int dx \phi(x)\sech(\epsilon x). \label{hep}
\eeq
The contamination of the relativistic modes into this projection is
\beq
\frac{\epsilon}{2}\int dx \pin{k}\phi_k\g_k(x,0)\sech(\epsilon x)=\pin{k}\phi_k J_k
\eeq
where
\beq
J_k=\frac{\epsilon}{2}\int dx \g_{k}(x,0) \sech(\epsilon x)=\frac{\pi\epsilon^2}{2k^2}\sech\left(\frac{\pi k}{2\epsilon}\right).
\eeq

The leading $\epsilon\sech$ term in $J_k$ vanishes, and this subleading $\epsilon^2\sech$ term will be affected by $O(\epsilon^3)$ and higher-order contributions to $\g_k$, which we have not computed.  Therefore, even the exponentially suppressed mixing seen here is an artifact of our truncation. While it is not clear that this exponential suppression occurs at all orders in the Fodor expansion, the projection Eq.~(\ref{hep}) of $\phi(x)$ onto $\hat\epsilon$ appears to work up to corrections of order $O(\epsilon^3)$.

In Ref.~\cite{quantosc} the operator $\pi_0$, whose eigenvalue is proportional to the oscillon's momentum, was defined by the projection
\beq
\pi_0=\frac{\epsilon^2}{\Omega^2}\int dx \pi(x) \tanh(\epsilon x)\sech (\epsilon x). \label{pi0}
\eeq
Inserting the relativistic modes Eq.~(\ref{reldec}) into $\pi(x)$ on the right-hand side, one finds a new contribution of
\beq
\frac{\epsilon^2}{\Omega^2}\pin{k}\pi_k I_k
\eeq
where 
\beq
I_k=\ok{}\left(1+\frac{4\epsilon^2}{k^2}\right)\int dx \g_{D,k}(x)\tanh(\epsilon x)\sech(\epsilon x)=-\frac{i\pi\omega_k}{k}\sech{\left(\frac{\pi k}{2\epsilon}\right)}.
\eeq
Here again the leading order term, proportional to $k\ok{}/\epsilon^2\sech(\pi k/(2\epsilon)$, vanished and this subleading term will receive contributions from higher orders in the Fodor expansion of $\g$ that we have not yet calculated. In conclusion, the old decomposition Eq.~(\ref{pi0}) continues to project onto the operator $\pi_0$ even given the relativistic corrections found in the present note.  One consequence of this is that its commutators reported in Ref.~\cite{quantosc} are unaffected by these relativistic corrections. 

More generally, the Fourier transform of $\sech^n{(\epsilon x)}$ is proportional to $\sech{(\pi k/2\epsilon)}$ for odd $n$ and to $\csch{(\pi k/2\epsilon)}$ for even $n$, up to polynomial factors in $k/\epsilon$, and is therefore suppressed as $e^{-\pi |k|/2\epsilon}$. For $n=2$, the complete overlap with the dual mode vanishes identically, while for full nonrelativistic modes such as Eq.~(3.7), any remaining overlap is exponentially suppressed and hence vanishes to all orders in $\epsilon/|k|$.

\subsection{Commutators}

As each $\phi_k$ is a linear combination of the $\phi(x)$ which commute with one another, the $\phi_k$ also commute with one another, and similarly so do the $\pi_k$
\beq
[\phi_{k_1},\phi_{k_2}]=[\pi_{k_1},\pi_{k_2}]=0.
\eeq
One may compute
\bea
\left[\phi_{k_1},\ok{2}{\pi_{k_2}}\right]&=&\frac{k_2^2}{k_2^2+4\epsilon^2}\int dx\int dy [\phi(x),\pi(y)]\g^*_{D,k_1}(x)\g^*_{k_2}(y)\\
&=&i\frac{k_2^2}{k_2^2+4\epsilon^2}\int dx \g^*_{D,k_1}(x)\g^*_{k_2}(x)=i\frac{k_2^2}{k_2^2+4\epsilon^2}\int dx \g^*_{D,k_1}(x)\g_{-k_2}(x)\nonumber\\
&=&\frac{k_2^2}{k_2^2+4\epsilon^2}2\pi i \delta(k_1+k_2).\nonumber
\eea

Let us define the operator
\beq
b_{-k}=\sqrt{\frac{\ok{}}{2}\left(1+4\frac{\epsilon^2}{k^2}\right)}\left(\phi_k+i\pi_k\right)
\eeq
so that
\beq
b^\dag_{k}=\sqrt{\frac{\ok{}}{2}\left(1+4\frac{\epsilon^2}{k^2}\right)}\left(\phi_k-i\pi_k\right).
\eeq
Then we may compute
\bea
[b_{-k_1},b_{-k_2}]&=&\frac{i}{2}\sqrt{\ok{1}\ok{2}\left(1+4\frac{\epsilon^2}{k_1^2}\right)\left(1+4\frac{\epsilon^2}{k_2^2}\right)}\left([\phi_{k_1},\pi_{k_2}]+[\pi_{k_1},\phi_{k_2}]\right)\label{bcom}
\\
&=&-\frac{1}{2}\sqrt{\ok{1}\ok{2}\left(1+4\frac{\epsilon^2}{k_1^2}\right)\left(1+4\frac{\epsilon^2}{k_2^2}\right)}
2\pi \delta(k_1+k_2)\left(\frac{1}{\ok{2}}\frac{k_2^2}{k_2^2+4\epsilon^2}-\frac{1}{\ok{1}}\frac{k_1^2}{k_2^1+4\epsilon^2}\right)\nonumber\\
&=&0 \nonumber
\eea
and similarly
\beq
[b^\dag_{k_1},b^\dag_{k_2}]=0.
\eeq
The commutation of the $b$ operators in Eq.~(\ref{bcom}) is essential to the quantization of the oscillon in \cite{quantosc}, because the ground state $\vac$ described there\footnote{The ground state is actually $\df\vac$ where $\df$ is a displacement operator.} is annihilated simultaneously by all $b_k$ operators.  Such a state only exists if the $b_k$ operators commute, but the commutation of the $b_k$ operators was only shown for nonrelativistic $k$, corresponding to those of Sec.~\ref{nrsez} of the present work.  The demonstration here that the relativistic modes also commute ties up an important loose end from that previous work.

Similarly, one finds
\beq
[b_{k_1},b^\dag_{k_2}]=\frac{i}{2}\sqrt{\ok{1}\ok{2}\left(1+4\frac{\epsilon^2}{k_1^2}\right)\left(1+4\frac{\epsilon^2}{k_2^2}\right)}
\left(-[\phi_{-k_1},\pi_{k_2}]+[\pi_{-k_1},\phi_{k_2}]\right)=2\pi\delta(k_1-k_2).
\eeq
Therefore, $b^\dag_k$ and $b_k$ satisfy the usual oscillator algebra, and so may be interpreted as creation and annihilation operators for Floquet modes in the quantum theory.  Note that, unlike the case of normal modes, the fact that Floquet mode creation and annihilation operators commute at distinct $k$ was not a consequence of the spectral theorem.  This is because the Floquet phase
\beq
e^{i\nu}=e^{-2\pi i \ok{}/\Omega}
\eeq
of distinct modes $k_1$ and $k_2$ agree if $\ok{2}-\ok{1}$ is an integral multiple of $\Omega$, and so one might expect such modes to be mixed.


One can easily show that
\beq
[\pi_0,\hat\epsilon]=[\phi_k,\hat\epsilon]=[\pi_0,\pi_k]=0.
\eeq
Using the commutators
\beq
[\pi_0,\phi_k]=-\frac{i\epsilon^2}{\Omega^2}\int dx \g_{D,k}^*(k)\sech(\epsilon x)\tanh(\epsilon x)= \frac{\pi \epsilon^2 k}{\Omega^2(k^2+4\epsilon^2)}\sech{\left(\frac{\pi k}{2\epsilon}\right)}
\eeq
\beq
[ 
\hat\epsilon,\pi_k]=\frac{i\pi \epsilon^2}{2\omega_k(k^2+4\epsilon^2)}\sech{\left(\frac{\pi k}{2\epsilon}\right)}
\nonumber
\eeq
we find
\beq
[\pi_0,b_{\pm k}]=\mp \frac{\pi \epsilon^2}{\Omega^2 }\sqrt{\frac{\omega_k}{2\left(k^2+4\epsilon^2\right)}}\sech{\left(\frac{\pi k}{2\epsilon}\right)}
\eeq
\beq
[\hat\epsilon
,b_{\pm k}]=-\frac{\pi \epsilon^2}{2k\sqrt{2\omega_k\left(k^2+4\epsilon^2\right)}}\sech{\left(\frac{\pi k }{2\epsilon}\right)},
\nonumber
\eeq
One can see that these commutators vanish as $e^{-\pi|k|/(2\epsilon)}$, and so do not contribute at any order in $\epsilon/k$.  This is again necessary for the existence of the oscillon ground state in Ref.~\cite{quantosc}, which is a simultaneous eigenstate of $\pi_0$, $\hat\epsilon$ and each operator $b_k$. Again, of course we have only expanded $\g_k$ to order $O(\epsilon^2/k^2)$ and so commutators may receive further corrections resulting from the $O(\epsilon^3)$ terms.

\section{Remarks}

In Ref.~\cite{quantosc}, a state in quantum field theory corresponding to an oscillon was constructed, and it was shown that it is periodic at leading order in the coupling and the Fodor expansion.  However, the state was defined as a simultaneous eigenstate of the boost operator $\pi_0$, the amplitude operator $\hat\epsilon$ and the continuum Floquet modes $b_k$, meaning that the state only exists if these operators commute.  It was shown that these operators commute; however, $b_k$ was only treated in the nonrelativistic regime $k\ll m$.  This demonstration was supplemented by a heuristic argument that the short ($k\gg \epsilon$) Floquet modes are plane waves when the oscillon amplitude is small, and so they should satisfy commuting copies of the oscillator algebra.  In the present note we have explicitly constructed the leading and subleading deviations of the continuum short Floquet modes from planewaves and we have shown that they lead only to commutators which are exponentially suppressed in the small parameter $\epsilon/|k|$.  Therefore, at the order at which we have worked in the $\epsilon/k$ expansion, the periodic state indeed exists.

The equations describing the perturbations of the small-amplitude \cite{spector,small97} Q-ball \cite{qball68,qball} are quite similar to those above describing the perturbations of the oscillon \cite{qpert08,qpert17,qpert18,qpert24,qpert24b,qpertnoi}.  Indeed, the Q-ball and the oscillon have many relations.  For example, a Q-ball may be seen as two oscillons out of phase \cite{swap14,o2q} and Q-ball solutions can be written using an artificial gauge redundancy and the gauge-invariance condition is in fact the oscillon solution \cite{renorm}.  Therefore we expect the results in the present note to carry over, essentially unchanged, to the Q-ball's continuum modes, allowing the full quantization of the Q-ball.  One-loop corrections to the quantum Q-ball have already been studied in \cite{q1loop,q12} while quantum Q-balls in the Hartree approximation have been studied in Refs.~\cite{har1,har2} and a saddle-point semiclassical approximation has been used in Ref.~\cite{qdec}.

Recently, a coherent state representing a boson star has been constructed in Ref.~\cite{saffin26}.  The one-loop divergences in the stress tensor \cite{erice} were treated by introducing ghost fields.  The approach of Refs.~\cite{quantosc,krakow}, in which the coherent state is squeezed so that all Floquet modes are in their ground states, may remove this divergence without the need to conjure ghosts, as it does in the case of the domain wall soliton \cite{noi3d}.

In this paper we believe that we have completed the calculation of the commutators of the basis of operators identified in Ref.~\cite{quantosc}.  With this in hand, the next step is to write the Hamiltonian in terms of these operators as in Ref.~\cite{krakow}.  This can be decomposed into a time-independent Floquet Hamiltonian, which describes dynamical processes that occur over entire periods, and a stroboscopic kick operator that describes events that are localized at time scales less than a period.  We do not yet know whether the Floquet Hamiltonian, like the soliton Hamiltonian that appears in the study of time-independent solitons, can be written as a sum of commuting copies of the harmonic oscillator plus nonlinear interactions which are suppressed by a power of the coupling.  

With these two Hamiltonians in hand, one can begin to understand the dynamics of quantum oscillons, attacking basic questions such as whether loop effects lead to faster spontaneous or induced radiation, reducing the oscillon's lifetime.  The lifetime of the oscillon determines whether electroweak oscillon collisions contribute significantly to baryogenesis, and also whether oscillons produced during preheating produce primordial black holes \cite{app5} or gravity waves \cite{app6,app7}.


\section*{Acknowledgement}

\noindent
This work was supported by the Higher Education and Science Committee of the Republic of Armenia (Research Project No.  24RL-1C047). BB is supported by the ANSO scholarship CAS.

\end{document}